\documentclass[letterpaper, 10pt]{IEEEtran}

\usepackage{amssymb}
\newtheorem{theo}{Theorem}
\newtheorem{cor}{Corollary}
\newtheorem{prop}{Proposition}

\newtheorem{ex}{Example}

\title{
On the complexity of computing the capacity of codes that avoid forbidden difference patterns}

\author{Vincent D. Blondel, Rapha\"el Jungers\thanks{V. Blondel and R. Jungers   are with the Department of Mathematical
Engineering, Universit\'e catholique de Louvain, Avenue Georges Lemaitre 4, B-1348 Louvain-la-Neuve, Belgium,
{\tt\small \{jungers, blondel\}@inma.ucl.ac.be}. Their work is supported by a grant ARC "Communauté française de
Belgique". Rapha\"el Jungers is a FNRS fellow (Belgian Fund for Scientific Research). }, and Vladimir Protasov%
\thanks{ V.Protasov is with the department of Mechanics and
Mathematics, Moscow State University, Vorobyovy Gory, Moscow, 119992, Russia, {\tt\small vladimir\_
protassov@yahoo.com}. His research is supported by the grant RFBR 05-01-00066 and by the grant 304.2003.1 for the
leading scientific schools. }}
\begin{document}

\maketitle 

\begin{abstract}
We consider questions related to the computation of the capacity of codes that avoid forbidden difference
patterns. The maximal number of $n$-bit sequences whose pairwise differences do not contain some given forbidden
difference patterns increases exponentially with $n$. The exponent is the capacity of the forbidden patterns,
which is given by the logarithm of the joint spectral radius of a set of matrices constructed from the forbidden
difference patterns. We provide a new family of bounds that allows for the approximation, in exponential time, of
the capacity with arbitrary high degree of accuracy. We also provide a polynomial time algorithm for the problem
of determining if the capacity of a set is positive, but we prove that the same problem becomes NP-hard when the
sets of forbidden patterns are defined over an extended set of symbols. Finally, we prove the existence of
extremal norms for the sets of matrices arising in the capacity computation.  This result makes it possible to
apply a specific (even though non polynomial) approximation algorithm.  We illustrate this fact by computing
exactly the capacity of codes that were only known approximately.

\end{abstract}


\section{Introduction}

In certain coding applications one is interested in binary codes whose elements avoid a set of forbidden patterns.
In order to minimize the error probability of some particular magnetic-recording systems, a more complicated
problem arises when it is desirable to find  code words whose \emph{differences} avoid forbidden patterns.

Let $\{0, 1\}^n$ denote the set of  words of length $n$ over $\{0, 1\}$ and let $u, v \in \{0, 1\}^n$. The
difference $u-v$ is a word of length $n$ over $\{-1, 0, +1\}$ (as a shorthand we shall use $\{-, 0, +\}$ instead
of $\{-1, 0, +1\}$). This difference is obtained from $u$ and $v$ by symbol-by-symbol subtraction so that, for
example, $0110 - 1011=-+0-$. Consider now a finite set $D$ of words over $\{-, 0, +\}$; we think of $D$ as a set
of \emph{forbidden difference patterns}. A set (or \emph{code}) $C \subseteq \{0, 1\}^n$ is said to \emph{avoid}
the set $D$ if none of the differences of words in $C$ contain a word from $D$ as subword, that is, none of the
differences $u-v$ with $u, v \in C$ can be written as $u-v=x d y$ for $d \in D$ and some (possibly empty) words
$x$ and $y$ over $\{-, 0, +\}$.

We are interested in the largest cardinality, which we denote by $\delta_n(D)$, of sets of words of length $n$
whose differences avoid the forbidden patterns in $D$. If the set $D$ is empty, then there are no forbidden
patterns and $ \delta_n(D)=2^{n}$. When $D$ is nonempty, then $ \delta_n(D)$ grows exponentially with the word
length $n$ and is asymptotically equal to $2^{\mbox{cap}(D) n}$ where the scalar $0 \leq \mbox{cap}(D) \leq 1$ is
the \emph{capacity} of the set $D$. The capacity is thus a measure of how constraining a set $D$ is; the smaller
the capacity, the more constraining the forbidden difference patterns are.

As an illustration consider the set of forbidden patterns $D=\{+-,++\}$. Differences between two words in
$C=\{u_{1}0u_{2}0 \cdots0 u_k :  u_{i}\in\{0,1\}\}$ will have a ``$0$" in any succession of two characters and
will therefore not contain any of the forbidden patterns.  From this it follows that $\delta_{n}\geq 2^{\lceil n/2
\rceil}$ and so $\mbox{cap}(D)\geq 1/2$. One can show in fact that $\mbox{cap}(D)= 1/2$. This follows from the
next proposition together with the simple observation that the capacity of the set $D=\{+-,++\}$ is identical to
the capacity of the set $D=\{+-,++,-+,--\}$.

 \begin{prop}\label{ex-cap-complet}
The capacity of the set $\{+,-\}^{m}$ is given by $({m-1})/{m}$.
\end{prop}
\begin{proof}
Let $C_{km}$ be a code of length $km$ avoiding $D$. In any given window of length $m$, the set of words appearing
cannot contain both $u$ and $\bar u$ (we use $\bar u$ to denote the word obtained by inverting the ones and the
zeros in $u$). This implies that there are at most $2^{m-1}$ different words in any given window of size $m$. Let
us now consider words in $C_{km}$ as a concatenation of $k$ words of length $m$. There are at most $ 2^{(m-1)k}$
words in $C_{km}$ and so $\mbox{cap}(D)\leq ({m-1})/{m}. $ Now consider the code
\begin{equation}
C_{km}=\{z_10z_20 \cdots 0z_k: z_i \in \{0, 1\}^{m-1}\}.
\end{equation}
 This code satisfies the constraints, and so the bound $(m-1)/m$ is reached.
\end{proof}

 The computation of the capacity is
not always that easy. As an example it is proved in \cite{moision01codes} that the capacity of
$\left\{+++\right\}$ is given by $\log_2((1+(19+3 \sqrt{33})^{1/3}+(19-3 \sqrt{33})^{1/3})/3)=.8791\ldots$ and the
same reference provides numerical bounds for the capacity of  $\left\{0+-+\right\}$ for which no explicit
expression is known.

The capacity of codes that avoid forbidden difference patterns was first introduced and studied by  Moision,
Orlitsky and Siegel. In \cite{moision01codes}, these authors provide explicit values for the capacity of
particular sets of forbidden patterns and they prove that, in general, the capacity of a forbidden set $D$ can be
obtained as the logarithm of the joint spectral radius of a set of matrices with binary entries. The joint
spectral radius, which we formally define below, is a quantity that quantifies the maximal asymptotic growth rate
of products of matrices taken from a set. This quantity is notoriously difficult to compute and approximate. It is
known in particular that the problem of computing, or even approximating, the joint spectral radius of two
matrices with binary entries is a problem that is NP-hard \cite{tsitsiklis97lyapunov} and that the problem of
determining if the joint spectral radius of two matrices with nonnegative entries is greater than one is
undecidable \cite{VB}. Moreover, the size of the matrices constructed in order to compute the capacity is not
polynomial in the size of the forbidden set $D$ : if the length of the forbidden words is $m$, the dimension of
the matrices is $2^{m-1}\times 2^{m-1}$.  Hence, even constructing the matrices is an operation that cannot be
performed in  polynomial time. However, as pointed out in \cite{moision01codes}, the matrices that arise in the
context of capacity computation have a particular structure and so these negative results do not rule out the
possibility for the capacity to be computable in polynomial time.

We provide several results in this paper, all are related to the capacity computation and its complexity.

We first provide new bounds that relate the capacity of a set of forbidden patterns $D$ with the values
$\delta_n(D)$, the maximum size of any code of length $n$ avoiding $D$. These bounds depend on parameters that
express the number and positions of zeros in the patterns of $D$. These new bounds allow us to compute the
capacity of any set to any given degree of accuracy by numerically evaluating $\delta_n (D)$ for increasing values
of $n$. The approximation algorithm resulting from these bounds has exponential growth but provides an a-priori
guaranteed precision, and so with this algorithm the computational effort required to compute the capacity to a
given degree of accuracy can be evaluated before the calculations are actually started. As an example, it follows
from the bounds we provide that the capacity of a set of forbidden pattern that does not contain any 0s can be
computed with an accuracy of 90$\%$ by evaluating $\delta_n(D)$ for $n=10$ (see Corollary \ref{cor2} below).

In a subsequent section, we provide explicit necessary and sufficient conditions for a set to have positive
capacity and we use this condition for producing a polynomial time algorithm that decides whether or not the
capacity of a set is positive.  As explained above, the capacity of a set is given by the logarithm of the joint
spectral radius of a set of matrices constructed from the forbidden patterns. Our polynomial time algorithm
therefore provides a procedure for checking whether or not the joint spectral radius of these matrices is larger
than one.  This problem is known to be NP-hard for general matrices.

We then consider the situation where in addition to the forbidden symbols $-, 0$ and $+$ the forbidden patterns in
$D$ may also include the symbol $\pm$, where $\pm$ stands for both the symbol $+$ and $-$. We prove that in this
case the problem of computing the capacity, or even determining if this capacity is positive, becomes NP-hard.

Finally, we show an algebraic property of the sets of matrices constructed in order to compute the capacity: there
exists always an extremal norm for this set.  This theoretical result makes it possible to apply specific
algorithms in order to compute the joint spectral radius.  These methods, although non-polynomial, can be more
efficient than general ones, and we use them for computing the capacity of some specific forbidden sets.

These results allow us to better delineate the capacity computation problems that are polynomial time solvable
from those that are not. We do however not provide in this paper an answer to the question, which was the original
motivation for the research reporter here, as to whether or not one can compute the capacity of sets of forbidden
patterns over $\{-, 0, +\}$ in polynomial time. This interesting question that was already raised in
\cite{moision01codes}, remains unsettled.


\section{Capacity and joint spectral radius} \label{section-jsr}

Let $D$ be a set of forbidden patterns over the alphabet $\{-, 0, +\}$ and consider for any $n \geq 1$ the largest
cardinality, denoted by $\delta_n(D)$, of sets of words of length $n$ whose pairwise differences avoid the
forbidden patterns in $D$. The capacity of $D$ is defined by
\begin{equation}\label{capdef}
\mbox{cap}(D)=\lim_{n\rightarrow \infty}\frac{\log_2 \delta_{n}(D)}{n}.
 \end{equation}

Moision et al. show in \cite{moision01codes} how to represent codes as products of matrices. Associated to any set
$D$ of forbidden patterns, they construct  a finite set $\Sigma(D)$ of matrices for which
 \begin{equation}\label{eq-deltan-rho}
\delta_{m-1+n}=\max_{}\{{\Vert A_{1} \dots A_{n}\Vert:A_{i}\in \Sigma(D)\}}.
 \end{equation}

In this expression, the matrix norm used is the sum of the absolute values of the matrix entries. The matrices
they construct are of dimension $2^{m-1}\times 2^{m-1}$ and have binary entries. For sake of conciseness, we do
not reproduce here the explicit construction of $\Sigma(D)$ but refer instead the interested reader to
\cite{moision01codes} for more details. Combining (\ref{eq-deltan-rho}) and (\ref{capdef}) we deduce that
\begin{eqnarray*}
\mbox{cap}(D) & = &\lim_{n\rightarrow \infty} \frac{\log_2 \delta_n(D)} {n}\\ & = &\lim_{n\rightarrow \infty}
\frac{\log_2 \delta_{m-1+n}(D)} {m-1+n}\\ & = &\lim_{n\rightarrow \infty}
\frac{\log_2 \max_{A_i \in \Sigma} \| A_{1} \dots A_{n}\|} {n}\\
 & = &  \log_2
\lim_{n\rightarrow \infty} \max_{A_i \in \Sigma} \| A_{1} \dots A_{n}\|^{1/n}
 \end{eqnarray*}

The quantity $\lim_{n\rightarrow \infty} \max_{A_i \in \Sigma} \| A_{1} \dots A_{n}\|^{1/n}$ appearing in the last
identity is a {\emph{joint spectral radius}}. For any compact set of matrices $A$, the joint spectral radius of
$A$ is defined by
$$\rho(A)=\limsup_{n\rightarrow \infty} \sup_{A_i \in A} \| A_{1}
\dots A_{n}\|^{1/n}.$$  Hence we have the fundamental relation :
$$\mbox{cap}(D) =\log_2 \rho(\Sigma(D)).$$  The joint spectral
radius of a set of matrices is a quantity that was introduced by Rota and Strang \cite{rota-strang} and that has
received intense research attention in the last decade. For more references on the joint spectral radius, consult
the survey \cite{blondel98survey}.


\section{Upper and lower bounds} \label{section-bounds}

In this section, we derive bounds that relate the capacity of a set $D$ with $\delta_n(D)$. Consider some set $D$
of forbidden patterns and  denote by $r_1$ (respectively $r_{2}$) the maximal $k$ for which $0^{k}$ is the prefix
(respectively suffix) of some pattern in $D$. No pattern in $D$ begins with more than $r_{1}$ zeros and no pattern
in $D$ ends with more than $r_{2}$ zeros. We also denote by $r$  the maximal number of consecutive zeros in any
pattern in $D$; obviously, $r \ge \max (r_1, r_2)$. In the next theorem we provide upper and lower bounds on the
capacity in terms of  $\delta_n(D)$. As $n$ increases these bounds converge to the same identical limit,
${\rm{cap}}(D)$, and so the bounds can be used to  approximate the capacity to any desired degree of accuracy.

\begin{theo}\label{th1}
For any $n \ge r_1 + r_2$ we have
 \begin{eqnarray}
   \frac{\log_2 \delta_{n}(D)-(r_1+r_2)}{n+r+1-(r_1+r_2)}
   \leq    {\rm{cap}} (D)
   \leq  \frac{\log_2 \delta_{n}(D)}{n}.
   \end{eqnarray}
\end{theo}
\begin{proof}
Let us first consider the upper bound.  The following equation is straightforward, given any positive integers
$k,n$, and any set of forbidden patterns $D$:
 \begin{equation}\nonumber
\delta_{kn}\leq \delta_{n}^{k}.
 \end{equation}
Indeed, considering any word of length $kn$ as the concatenation of $k$ subwords of length $n$, for each of these
subwords we have at most $\delta_{n}$ possibilities.  Taking the $\frac{1}{kn}$th power of both sides of this
inequality and taking the limit $k\rightarrow \infty$, we obtain :
 \begin{eqnarray}\nonumber
2^{\rm{cap} (D)}&\leq &\delta_{n}^{1/n}\\\nonumber
 {\rho}^{n}&\leq & \delta_{n}.
 \end{eqnarray}
Now let us consider the lower bound.  The optimal code of length $n$ contains at least $\lceil 2^{-r_1 -
r_2}\delta_{n} (D)\rceil$ words that coincide in the first $r_1$ bits and in the last $r_2$ bits (because there
are in total $2^{r_1 + r_2}$ different words of length $r_1 + r_2$). Denote the set of strings of all these words
from $(r_1+1)$st bit to $(n-r_2)$th bit by $C'$. This set contains at least $\lceil 2^{-r_1 -
r_2}\delta_{n}(D)\rceil$ different words of length $n-r_1 - r_2$. Then for any $l \ge 1$ the code $$ C = \bigl\{
u_{1}0^{r+1}u_{2}0^{r+1} \cdots 0^{r+1}u_{l}0^{r+1}\, , \ u_k \in C',\, k = 1, \ldots , l\bigr\}
$$ avoids $D$. The cardinality of this code is at least $\lceil
2^{-r_1 - r_2}\delta_{n}(D)\rceil^{l}$ and the length of its words is $N = l(n-~r_1 - r_2 + r + 1)$. Therefore,
for any $l$ we have
$$ \delta_{N} (D) \ \ge \ \lceil 2^{-r_1 -
r_2}\delta_{n}(D)\rceil^{l}. $$ Taking the power $1/N$ of both sides of this inequality, we get $$ \Bigl[
\delta_{N} (D) \Bigr]^{1/N} \ \ge \ \lceil 2^{-r_1 - r_2}\delta_{n}(D)\rceil^{1/(n-r_1 - r_2 + r + 1)}, $$ which
as $N \to \infty$ yields $$
 \rho  \ \ge \ \lceil 2^{-r_1 - r_2}\delta_{n}(D)\rceil^{1/(n-r_1 - r_2 + r
+ 1)}. $$ Now after elementary simplifications we arrive at the lower bound on ${\rm{cap}} (D)$.
\end{proof}

These bounds allow to design an algorithm in order to compute the capacity only by evaluating the successive
values $\delta_n$.  For an efficient (although non-polynomial) method of computation of $\delta_n$, see
\cite{moision-joint-constraints}.

Both bounds in this theorem are sharp in the sense that they are both attained for particular sets $D$. The upper
bound is attained for the set $D=\emptyset$ and the lower bound is attained, for instance, for the set $D =
\{0^{m-1}+ \}$. Indeed, in this case $ r = r_1 = m-1, r_2 = 0$ and ${\rm cap} (D)=0$, while $\delta_{n}=2^{m-1}$
for $n\ge m-1$.  Here is a direct proof of this equality, drawn from \cite{moision01codes}: Clearly, for all
$n>m-1$, we can construct a code of size $\delta_{n}=2^{m-1}$. It happens that for any given length $n$ this size
is maximum. Otherwise, there must be  two \emph{different} words $u$ and $v$ whose prefixes of length $k$
coincide. In order to avoid the forbidden pattern, the $k+1$-th symbols must also be equal, and so on. But then
both words are equal, and we have reached a contradiction.

\begin{cor}\label{cor1}
Let $D$ be given and let $r, r_1$ and $r_2$ be defined as above. Then

$ {\rm cap} (D)\leq \frac{\log_2 \delta_{n}}{n} \leq\frac{1}{n} \max (r_1 + r_2, r+1) + {\rm cap} (D)$.

\end{cor}

\begin{proof}
Simply use the fact that the capacity is always less than one.
\end{proof}

We may specialize these general bounds to sets of particular interest.
\begin{cor}\label{cor2}
Let $D$ be given and let $r, r_1$ and $r_2$ be defined as above. Then
\begin{enumerate}
 \item \label{c6} If ${\rm{cap}} (D) = 0$ the
size of any code avoiding $D$ is bounded above by the constant $2^{r_1 + r_2}$. \item \label{c5} If the patterns
in $D$ contain no zeros, then $$  n \; {\rm cap} (D) \le \log_2 \delta_n (D) \le (n+1) \; {\rm cap} (D).$$ \item
\label{c4} If none of the patterns in $D$ start or end with a zero, then $  n \; {\rm cap} (D) \le \log_2 \delta_n
(D) \le (n+r+1) \; {\rm cap} (D)$.

\end{enumerate}
\end{cor}

\begin{ex}\label{pzpzpz}
   We do not know the capacity of the set $D=\{+0+0+\}$ with a good precision,
but by Corollary \ref{prop-un} below, we know that it is positive. In this case $r_1 = r_2 =0$, $r=1$, and hence,
$$
  \rho^{n}\ \leq \delta_{n}\ \leq  4  \rho^{n}.\\
$$
\end{ex}
\begin{ex}\label{ppm1}
Consider the set $D=\{++-\}$.
 A approximate computation of the joint spectral radius leads to $  \rho  = 2^{0.8113\dots} < 1.755 $, and thus, by Corollary \ref{cor2},
 $$
2^{0.8113n}\ \leq \delta_{n} \  \leq \, 1.755\  2^{0.8113n}. $$
\end{ex}


\section{Positive capacity can be decided in  polynomial time} \label{section-complexity}

As previously seen by a direct argument, the capacity of the set $\{0^{m-1}+\}$ is equal to zero. In this section
we provide a systematic way of deciding when the capacity of a set is equal to zero. We first provide a simple
positivity criterion that can be verified in finite time and then exploit this criterion for producing a
positivity checking algorithm that runs in polynomial time. In the sequel we shall use the notation $-D$ to denote
the set of elements that are the opposites to the elements of  $D$, for example if $D=\{-+0, 0--\}$ then
$-D=\{+-0, 0++\}$.

\begin{theo}\label{theo-criterion}
Let  $D$ be a set of forbidden patterns of lengths at most $m$. Then ${\rm{cap}} (D) >  0$ if and only if there
exists a word on the alphabet $\{+, -,0\} $ that does not contain any word of $D \cup -D$ as subword and that has
a prefix $0^m$ and a suffix $+0^{m-1}$.
\end{theo}
\begin{proof} If the capacity is positive, then for sufficiently
large $n$ there is a code avoiding $D$ of size  $\ge 2^{2m-1}+1$. This code has at least two words $u,v$ with the
same $m$-bit prefix and $(m-1)$-bit suffix (because there are in total $2^{2m-1}$ different words of the length
$2m-1$). Taking the difference $u-v$ and removing, if necessary, several last zeros we get an admissible string
with a prefix $0^m$ and a suffix $\pm 0^{m-1}$. If the first bit in the suffix is $+$, then the proof is
completed, if it is $-$, then we apply the same reasoning to $v-u$.

Conversely, suppose there exists a feasible string $d$ of $+, -$ and $0$ of some length $n$. Clearly, $n \ge 2m$.
Let $u_1, u_0 \in \{0, 1\}^{n-2m+1}$ be binary words, for which $0^{m}(u_1- u_0)0^{m-1}= d$. Then for any $l \ge
1$ the code
\begin{equation}\label{codespec}
C = \bigl\{ u_{i_1}0^{m-1} \cdots u_{i_l}0^{m-1}\, , \ i_k \in \{0, 1\},\, k = 1, \ldots , l\bigr\}
\end{equation}
avoids $D$. The cardinality of this code is $2^{l}$.  Hence $\delta_{n}(D)$ is unbounded and we conclude from
Corollary \ref{c6} that the capacity is positive.
\end{proof}

\begin{cor}\label{cor-avec-zero} If every word in $D$ contains at
least two nonzero symbols, then $\mbox{cap}(D)>0$.
 \end{cor}
\begin{proof} For any such set the word $d = 0^m + 0^{m-1}$
is admissible, and by Theorem \ref{theo-criterion} the capacity is positive.
\end{proof}

\begin{cor}\label{prop-un}
If $D$ consists of one forbidden pattern $p$ of length $m$, then its capacity is zero if and only if $p$ has at
least $m-1$ consecutive zeros.
 \end{cor}

\begin{proof} If a pattern $p$ is $0^m$ or $+0^{m-1}$, then obviously
there are no admissible strings, and by Theorem \ref{theo-criterion} the capacity is zero. The same holds for
$-0^{m-1}$, since this is the negation of $+0^{m-1}$ and for $0^{m-1}\pm$ because of the symmetry. In all the
other cases the admissible string exists and so $\rm{cap} (D) > 0$. Indeed, if $p$ has a unique nonzero bit, then
the word $d = 0^m + + 0^{m-1}$ is admissible, if it has at least two nonzero bits, then the proof follows from
Corollary \ref{cor-avec-zero}.
\end{proof}

We now prove the polynomial-time solvability of the problem of determining whether the capacity of a set $D$ is
positive.  The proof is constructive and is based on the so-called \emph{Aho-Corasick} automaton that checks
whether a given text contains as a subsequence a pattern out of a given set \cite{algo-aho-corasick}.  Let $P$ be
a given set of patterns. The transition graph of the Aho-Corasick automaton for the set $P$ is defined as follows
: First, construct the \emph{retrieval tree}, or trie, of the set $P$.  The trie of $P$ is the directed tree of
which each vertex has a label representing a prefix of a pattern in $P$, and all prefixes are represented,
including the patterns themselves. The root of the tree has the empty string as label. Edges have a label too,
which is a symbol of the used alphabet. There is an edge labeled with the symbol $a$ from a vertex $s$ to a vertex
$t$ if $t$ is the concatenation $sa$.\\ In order to have an automaton, we complete the trie by adding edges so
that for each vertex $s$, and each symbol $a$, there is an edge labeled $a$ leaving $s$.  This edge points to the
vertex of the trie of which the label is the longest suffix of the concatenation $sa$. Note that this vertex can
be the root (that is, the empty string) if no vertex in the trie is a suffix of $sa$. Finally, the accepting
states of the automaton are the vertices whose labels are patterns of $P$.  This automaton accepts words that
contain a pattern in $P$ and halts whenever this pattern is a suffix of the entered
text.\\

If $0^{k}\in D$ or $+0^{k}\in D,\, k\leq m$, then, by Theorem \ref{theo-criterion}, $\mbox{cap}(D)=0$. If this is
not the case, we construct the graph of the automaton of Aho-Corasick for the set $P=D\cup (-D)\cup \{+0^{m-1}\}$.
We then remove any vertex labeled with a pattern in $P$ ({i.e.}, a state reached when a suffix of the text entered
is in the set $P$) except the vertex labeled  $\{+0^{m-1}\}$. The size of the constructed graph is polynomial in
the size and the number of the forbidden patterns. Let us now denote $q_{0^{m}}$ the state reached after entering
the word $0^{m}$.  This state is well defined since $0^{m}$ doesn't contain any forbidden pattern, and hence no
state reached after entering any prefix of the string $0^{m}$ was removed from the primary automaton.  We also
denote $q_{+0^{m-1}}$ the state corresponding to the suffix $+0^{m-1}$ for the entered text ({i.e.} the accepting
state corresponding to the pattern $+0^{m-1}$ in the Aho-Corasick automaton).  We have the following criterion for
zero-capacity:
%
\begin{theo}\label{th4}
The capacity of a set $D$ is positive if and only if there is a path from $q_{0^{m}}$ to $q_{+0^{m-1}}$ in the
graph constructed above.\end{theo}

\begin{proof}
If $\rm{cap} (D)>0$, by Theorem \ref{theo-criterion}, there exists a word $d$, beginning with $m$ zeros, and
ending with $+0^{m-1}$, that avoids $D\cup -D$. Hence, entering this word in the automaton, the finite state will
be (well defined and will be) the vertex labeled $+0^{m-1}$, because the vertices removed from the original
automaton of Aho-Corasick do not make any problem, since we do not reach the vertices labeled with forbidden
patterns.\\ On the other hand, a path in the constructed graph represents an acceptable word, since it doesn't
pass through any removed vertex, and hence no suffix of any prefix of this word will be in the forbidden set.\\
Moreover, a shortest path will give the shortest acceptable word, since the length of the path is equal to the
length of the represented word.
\end{proof}

\begin{cor}\label{cor-P}
The problem of determining whether or not the capacity of a given set of forbidden patterns is positive can be
solved in polynomial time.\end{cor}

\begin{proof}
Aho shows in \cite{algo-aho-corasick} that the automaton is constructible in polynomial time.  The determination
of the state $q_{0^{m}}$ and the computation of the shortest path are obviously polynomially feasible.
\end{proof}

\begin{cor}\label{c7}
If for a set $D$ of forbidden patterns there are admissible words, then the length of a shortest admissible word
does not exceed $2M+2m$, where $m$ is the maximal length of all patterns in $D$ and $M$ is the sum of the lengths
of each forbidden pattern.
\end{cor}
\begin{proof}
The number of vertices of the graph does not exceed $2M+m+1$. Indeed, for each pattern of length $l$ in $D\cup -D$
we add to the automaton at most $l$ states, since there are no more than $l$ prefixes of this pattern. We still
add the pattern $ \{+0^{m-1}\}$ (maximum $m$ new states), and the root. If there is a path connecting two given
vertices, this path can be chosen so that its length (in terms of number of vertices) will not exceed the total
number of vertices (if it does not pass through the same vertex twice).
  Every edge of this path adds
one bit to the admissible string. The initial length of the string is $m$ (we start from $0^m$), therefore the
total length of the admissible word is at most $2M+2m$.
\end{proof}

\begin{prop}\label{prop3}
If the capacity is positive, then ${\rm{cap}}(D)>{1}/({2M+m})$, where $M$ is the total number of characters in $D$
and $m$ is their maximal length.
\end{prop}
\begin{proof}
If $\rm{cap}(D) > 0$, then there is an admissible string of length $n \le 2M+2m$ (Corollary \ref{c7}). Consider
the code given by equation (\ref{codespec}). Its size is $2^{l}$ and the length of its words is at most

$$ N_l \ = \ l (2M+2m-m) =l \bigl ( 2M+m \bigr ) $$ Therefore $$
\begin{array}{c}
\rm{cap}(D) \ = \ \lim\limits_{l \to \infty}\frac{\log_2
\delta_{N_l}}{N_l}\\
 \ \ge \ \lim\limits_{l \to \infty}\frac{\log_2 2^l}{l\bigl (2M+m\bigr)}
 \ = \ \frac{1}{2M+m}.
\end{array}
$$
\end{proof}


\section{Positive capacity is NP-hard for extended sets}

We now consider the situation where forbidden patterns are allowed to contain the $\pm$ symbol. The symbol $\pm$
is to be understood in the following sense: whenever it occurs in a forbidden pattern, both the occurrences of $+$
and of $-$ are forbidden at that particular location. So, for example, avoiding the forbidden set $\{0\pm + \pm
\}$ is equivalent to avoiding the set $\{0+++, 0++-, 0-++, 0-+-\}$. All results obtained for forbidden patterns
over $\{-, 0, +\}$ have therefore their natural counterparts in the situation where the forbidden patterns are
defined over the alphabet $\{-, 0, +, \pm\}$. In particular, the results of Section 3 do transfer \emph{verbatim}
and the bounds derived in Theorem 1 are valid exactly as stated there. We now prove a complexity result of
capacity computation in this set-up.

\begin{theo} \label{tun}
The problem of determining if the capacity of a set of forbidden patterns over $\{0,+,-,\pm\}$ is equal to zero is
NP-hard.
\end{theo}
\begin{proof}
The proof proceeds by reduction from the Not-All-Equal 3SAT problem that is known to be NP-complete (see
\cite{GJ-computers-igt}). In the Not-All-Equal 3SAT problem, we are given $m$ binary variables $x_{1},\dots,x_{m}$
and $n$
 clauses that each contain three literals (a literal can be a variable or its
negation), and we
 search a truth assignment for the variables such that each clause has at least
one
 true literal and one false literal. \\
Suppose that we are given a set of clauses. We construct a set of forbidden patterns $D$ such that $\mbox{cap} (D)
>0$ if and only if  the instance of Not-All-Equal 3SAT has a solution. The first part of $D$ is given by:
\begin{equation}
\{(0 \pm 0),(0 \pm \pm 0),\dots, ( 0 \pm^{m-1}0)\}.
 \end{equation}
 Words over $\{-, 0, +\}$ that
avoid these patterns are exactly those words for which any two consecutive zeros are either adjacent or have at
least $m$ symbols $+$ or $-$ between them. We use these $m$ symbols as a way of encoding possible truth
assignments for the variables.

We then  add to $D$ two patterns for every clause. These patterns are of length $m$ and are entirely composed of
symbols $\pm$, except for the positions corresponding to the three variables of the clause, which we set to $+$
if the clause contains the variable itself, or to $-$  if the clause contains the negation of the variable. We
also add the opposite of this pattern; this last pattern is not necessary for the proof but preserves the symmetry
of the construction.

For example, if the instance of Not-All-Equal 3SAT consists of the two clauses $(x_1, \bar{x}_3, x_{4})$ and
$(\bar{x}_2, {x}_4, x_{5})$, the corresponding set $D$ will be $D=\{(0 \pm 0),(0\pm \pm 0),(0\pm \pm \pm 0), (0\pm
\pm \pm \pm 0),(+\pm - + \pm),(-\pm
+ - \pm), (\pm - \pm + +), (\pm + \pm - - ) \}$.\\
Such a set $D$ has always a length polynomial in the number of clauses and the number of variables.\\ We now prove
that there is a solution to the instance of Not-All-Equal 3SAT if and only if $\mbox{cap}(D)> 0$. First, suppose
that there exists a satisfying truth assignment for $x$ and  denote it by $\{\omega_{1},\dots,\omega_{m}\}$.
Associated to any $k \geq 1$ we construct a code of length $k(m+1)$ containing $2^k$ words as follows:
 $$C_{k(m+1)}= \{0  \omega 0 \omega 0 \omega 0 \cdots 0 \omega 0
 \omega,
 0  \omega 0 \omega 0 \omega 0 \cdots 0 \omega 0 \bar \omega,$$
 $$0  \omega 0 \omega 0 \omega 0\cdots 0 \bar \omega 0 \omega,
 \ldots,
 0  \bar \omega 0 \bar \omega 0 \bar \omega 0 \cdots 0 \bar \omega 0 \bar
 \omega\},$$
 where $\omega=\omega_1 \cdots \omega_m$.

Any difference between two words in this code is a word of the form $0z_10z_20 \cdots 0 z_k$ where for every $1
\leq i \leq k$, $z_i$ is either a sequence of $m$ 0's or a word of length $m$ over $\{-,+\}$. Because $\omega$
satisfies the instance of Not-All-Equal 3SAT, these words avoid the set $D$ constructed above. Moreover, the
cardinality of $C_{k(m+1)}$ is $2^{k}$ and hence
\begin{equation}
\mbox{cap}(D)\geq \lim_{k\rightarrow \infty} \log_{2} 2^{\frac{k}{k(m+1)}}=\frac{1}{m+1}>0.
\end{equation}

For the reverse implication, assume now that $\mbox{cap}(D)>0$. The capacity is positive, and so one can find two
words whose differences contain a 0 and a $+$. But then since this difference must avoid the first part of the
forbidden pattern, for a code $C$ large enough, there must exist two words in the code whose difference contains a
word over $\{-, +\}$ of length $m$. But this sequence avoids also the second part of $D$, and thus it represents
an acceptable solution to our instance of Not-All-Equal 3SAT.
 \end{proof}

Note that a similar proof can be given if we replace the symbol ``$\pm$" in the statement of the theorem by a
symbol that represents either $+$, $-$, or $0$.


\section{Extremal norms and computing of the capacity} \label{section-extremal}

A classical way to estimate a joint spectral radius consists of computing successive upper bounds on it by
applying the following well known inequality
 \begin{eqnarray}\label{ineq-jsr}
\nonumber \rho^{n}(\Sigma) \leq \max{\{\Vert A_{1} \dots  A_{n} \Vert : A_{i} \in \Sigma\}}
 \end{eqnarray}
that holds for any norm, and any length $n$ of the products.  One could then hope that for a well-chosen norm, the
joint spectral radius would be already obtained for $n=1$, that is, for the set of matrices itself. This is the
concept of extremal norm, which we now define properly : A norm $\|\cdot \|$ in $\mathbb R^d$ is called extremal
for a family of operators $A_1, \ldots , A_r$ if $\|A_i\| \le   \rho (A_1, \ldots, A_r)$ for all $i = 1, \ldots ,
r$. The unit ball $M$ of this norm is called the extremal convex body.

The notion {\em extremal} is justified by the fact that $ \rho$ is the smallest possible value such that the norms
of all the operators  $A_1, \ldots , A_r$ do not exceed this value. The above inequality, which holds for any
norm, for extremal norms becomes an equality (for all $n \ge 1$). If $M$ is the unit ball corresponding to an
extremal norm (an \emph{extremal body}), $A_iM \subset
  \rho M$. On the other hand, any convex body $M$ (convex
compact with nonempty interior centrally symmetric with respect to the origin) possessing this property generates
an extremal norm. It suffices to take the Minkowski norm defined by this body: $\|x\| = \inf \{\lambda > 0,\
\frac{1}{\lambda } x \in M\}$. Thus, there is a natural one-to-one correspondence between extremal norms and
extremal bodies.

The existence of an extremal norm can simplify many problems related to the joint spectral radius, see
\cite{barabanov}, \cite{protasov1} and \cite{protasov2} for details. However, not every set of matrices possesses
an extremal norm. The corresponding counterexamples are simple and well-known. Sufficient conditions for the
existence of an extremal norm can be found in \cite{barabanov}, \cite{protasov1}, \cite{protasov3}. For the
matrices arising in the context of the capacity computation, however, these conditions are not always satisfied.
Nevertheless, it turns out that in the case of capacity computation, the matrices do in fact always possess an
extremal norm.

\begin{theo}\label{th2}
For any set $D$ of forbidden patterns the set $\Sigma (D)$ possesses an extremal norm.
\end{theo}
\begin{proof} Let $\Sigma (D)$ be the set of matrices
corresponding to $D$. For a given point $x \ge 0$ let $$ \mathcal O (x)\ = \ \Bigl\{  \rho^{-n}A_{1}\dots
A_{n}x,\quad  A_{i}\in \Sigma (D)\ n \ge 0\Bigr\} $$ be the normalized orbit of the point $x$ under the action of
all possible products of the operators in $\Sigma$. The product of length zero is defined as the identity
operator, so the set $\mathcal O (x)$ contains $x$. Now define a set $M$ as follows $$ M \ = \ \rm{Conv}\,
\Bigl\{\mathcal{O}(\emph{e}_j) , \, \mathcal O (-\emph{e}_j),\ j = 1, \ldots , 2^{m-1} \Bigr\}, $$ where $e_1,
\ldots , e_{2^{m-1}}$ are the canonical basis vectors in $\mathbb R^{2^{m-1}}$, and $\rm{Conv}$ denotes the convex
hull. The set $M$ is obviously convex, centrally symmetric with respect to the origin, and possesses a nonempty
interior (because it contains the cross-polytope with the vertices $\{\pm e_j , \ j = 1, \ldots , 2^{m-1}\}$ whose
interior is nonempty).  Moreover, Theorem \ref{th1} implies that the set $\mathcal O (x)$ is bounded for any $x$,
therefore $M$ is bounded. So $M$ is a convex body that possesses the property $A_iM \subset   \rho M,\ i = 1,
\ldots , r$. Therefore it generates an extremal norm, and the theorem follows.

\end{proof}
 The very existence of an extremal norm for a set of
matrices makes it possible to apply a geometric algorithm for computing a joint spectral radius with a given
relative precision $\varepsilon$. We now briefly describe this algorithm; for all technical details we refer the
reader to \cite{protasov1}. For the sake of simplicity we consider the case of two matrices, the case of an
arbitrary number of matrices is treated in the same way.
\smallskip

{\em The algorithm.} Suppose operators $A_0, A_1$ acting in $\mathbb R^d$ possess an extremal norm; one needs to
find a number $\rho^*$ such that $\frac{\bigl|\rho^* - \rho \bigr|}{\rho } \, < \, \varepsilon , $ where $\,
\varepsilon >0$ is a given accuracy. Consider a sequence of convex polytopes $\{P_k\}$ produced as follows. $ P_0
= \Bigl\{(x_1, \ldots , x_d)\in \mathbb R^d ,\quad \sum |x_i| \le 1 \Bigr\} $ is a cross-polytope. For any $k \ge
0$ the polytope $P_{k+1}$ is an arbitrary polytope possessing the following properties:  it is symmetric with
respect to the origin, has at most $q(\varepsilon ) = C_d\, \varepsilon^{\frac{1-d}{2}}$ vertices, where $C_d$ is
an effective constant depending only on $d$, and $\, (1-\varepsilon)\bar A P_k \, \subset \, P_{k+1} \, \subset
\, \bar A P_k$, where $\bar A X = {\rm Conv \{A_0X, A_1X\}}$.

After $N = \Bigl[\frac{3\sqrt{d}\ln \frac{c_2}{c_1} }{\varepsilon }\Bigr]$ steps the algorithm terminates. The
value ${\rho^* = \bigl(v_{N+1} \bigr)^{1/(N+1)}}$ gives the desirable approximation of the joint spectral radius.
Here $v_k$
 is the biggest distance from the origin to the vertices of
the polytope $P_{k}$, $c_1, c_2$ are lower and upper bounds of the values $\, \{ \rho^{-k}\cdot {\rho}_{k}(\Sigma
,\Vert \cdot \Vert),\ k \in \mathbb N\}$. For the adjacency matrices the values $c_1, c_2$ can be taken directly
from Theorem \ref{th1}.

Each step requires us taking the convex hull of two polytopes having at most $q(\varepsilon)$ vertices and
requires the approximation of one polytope with $2q(\varepsilon)$ vertices
 by a polytope with $q(\varepsilon)$ vertices with accuracy
 $\varepsilon$. Both operations
 are known to be polynomial w.r.t. $\frac{1}{\varepsilon}$ \cite{protasov1}
 (the dimension $d$ is fixed). The computational complexity of this algorithm
is $ C \cdot\varepsilon^{-\frac{d+1}{2}} $, where $C$ is some constant and $d= 2^{m-1}$. Therefore the algorithm
is applicable for small values of $m$, say, for $m \le 6$.

The complexity of this algorithm is exponential with respect to $m$, as the one proposed in Section
\ref{section-bounds} that approximates the capacity by successive estimations of $\delta_n$. The advantages of one
algorithm over the other  appears in numerical computation of the capacity. In many cases the approximation of
invariant bodies by polytopes can lead to the exact value of the joint spectral radius. Suppose that by numerical
observations we conjecture that $\rho$ is attained by some product $\Pi_n = A_{i_1}\ldots A_{i_n}$, \emph{i.e.} $
\rho = \rho(\Pi_n)^{1/n}$. If during the calculations we find a polytope $P$ such that  $AP \subset
\rho(\Pi_n)^{1/n} P$, then it occurs that $\rho = \rho(\Pi_n)^{1/n}$. As the polytope $P$ we take $P= P_k={\rm
Conv }\{\bar A^jv\, , \, \bar A^jv\, , \ j=0, \ldots , k\}$ for some integer $k$, where $v$ is the eigenvector of
$\Pi_n$ corresponding to the largest by modulo eigenvalue (we assume that this is real and unique).

Let us  illustrate this method by  computing the exact values of the capacity for several codes. In Examples
\ref{ex1} and \ref{ex2} we find the values of capacities that were approximated in \cite{moision01codes}. Example
\ref{ex3} deals with a code with $m=4$.
\begin{ex}\label{ex1}
$\mbox{cap}(\{0++\})=\log_2 \rho (A_0) = \log_2 \bigl(\frac{\sqrt{5}+1}{2}\bigr)=0.69424191\ldots.$ The
eigenvector is $v=(2, \sqrt{5}-1, 2, \sqrt{5}-1)^T$. The algorithm terminates after five steps, the polytope $P =
P_5$  has 32 vertices.
\end{ex}
\begin{ex}\label{ex2}
$\mbox{cap}(\{0+-\})= \log_2 \rho (A_0) =\log_2 \bigl(\frac{\sqrt{5}+1}{2}\bigr).$ The algorithm terminates after
four steps, $v=(2, \sqrt{5}-1,  \sqrt{5}-1,2)^T, \ P = P_4$, the polytope has 40 vertices.
\end{ex}
\begin{ex}\label{ex3}
$\mbox{cap}(\{+++-\})= \log_2 \bigl(\frac{\sqrt{3+2\sqrt{5}}+1}{2}\bigr)= \log_2 \sqrt{\rho (A_0 A_1)}=
0.90053676\ldots.$  The algorithm terminates after eleven steps, the polytope $P = P_{11} $ has 528 vertices.
\end{ex}

As illustrated in many applications it is quite often the case that the joint spectral radius is attained by some
finite product. We say in these cases that the set of matrices possess the {\it finiteness property.}  It was
conjectured that all sets of matrices have the finiteness property : this is the well known finiteness conjecture
which has been disproved in \cite{cfbousch}, \cite{cfblondel}, and \cite{cfkoz}. Nevertheless, we conjecture here
that the sets of matrices constructed in order to compute a capacity do always possess the finiteness property.
Numerical results in \cite{moision01codes}, \cite{master-thesis}, and in this paper seem to support this
conjecture.

 \section{Conclusion}

One  way to compute the capacity of a set of forbidden patterns is to compute the joint spectral radius of a set
of matrices.  In practice, this leads to a double difficulty : first, the size of the matrices is exponential in
the size of the set of forbidden patterns, and second, the joint spectral radius is in general NP-hard to compute.
Actually, it is even NP-hard to decide whether a joint spectral radius of two matrices is greater than one. We
show in this paper that the simpler problem of checking the positivity of the capacity of a set defined on
$\{+,-,0\}$ is polynomially decidable but that the same problem becomes NP-hard when defined over the alphabet
$\{+,-,0, \pm \}$. We also provide bounds that allow faster computation of the capacity.  Finally we prove the
existence of extremal norms for the sets of matrices arising in the capacity computation and present a geometrical
algorithm for capacity computation which we illustrate with several numerical examples. Even if this latter result
allows to use algorithms that have proved to be quite efficient in practice, we should keep in mind that the
approach that consists of computing the joint spectral radius of the matrices defined in \cite{moision01codes}
cannot lead to a polynomial algorithm to compute the capacity because of the exponential size of the matrices.

\section*{Acknowledgment} We would like to thank Noga Alon and Alexander Razborov, both from the Institute for
Advanced Study (Princeton, USA), for providing an initial  proof of a result that is essentially equivalent to the
statement of Theorem 3. The proof presented here is different and is based on the Aho-Corasick automaton. We also
express our thanks to a student of Moscow State University, E.Shatokhin, for implementing the algorithm introduced
in this paper and for the numerical computation of the Examples \ref{ex1} to \ref{ex3}.

 \end{document}